\documentclass{article} 
\usepackage{iclr2021_conference,times}

\usepackage{amsmath,amsfonts,bm}

\def\eqref#1{equation~\ref{#1}}

\def\1{\bm{1}}

\DeclareMathAlphabet{\mathsfit}{\encodingdefault}{\sfdefault}{m}{sl}
\SetMathAlphabet{\mathsfit}{bold}{\encodingdefault}{\sfdefault}{bx}{n}

\usepackage{subcaption}
\usepackage{graphicx}
\usepackage{float} 
\usepackage{hyperref}
\usepackage{url}

\title{COVID-19 Detection from Chest X-ray Images using Imprinted Weights Approach}

\author{Jianxing Zhang, Pengcheng Xi, Ashkan Ebadi, Hilda Azimi \& St\'ephane Tremblay \\
Digital Technologies Research Centre, National Research Council of Canada, Canada\\
\texttt{\{firstname.lastname\}@nrc-cnrc.gc.ca} \\
\And
Alexander Wong \\
Department of Systems Design Engineering, University of Waterloo, Ontario, Canada \\
\texttt{alexander.wong@uwaterloo.ca} \\
}

\iclrfinalcopy 
\begin{document}

\maketitle

\begin{abstract}
The COVID-19 pandemic has had devastating effects on the well-being of the global population. The pandemic has been so prominent partly due to the high infection rate of the virus and its variants. In response, one of the most effective ways to stop infection is rapid diagnosis. The main-stream screening method, reverse transcription-polymerase chain reaction (RT-PCR), is time-consuming, laborious and in short supply. Chest radiography is an alternative screening method for the COVID-19 and computer-aided diagnosis (CAD) has proven to be a viable solution at low cost and with fast speed; however, one of the challenges in training the CAD models is the limited number of training data, especially at the onset of the pandemic. This becomes outstanding precisely when the quick and cheap type of diagnosis is critically needed for flattening the infection curve. To address this challenge, we propose the use of a low-shot learning approach named imprinted weights, taking advantage of the abundance of samples from known illnesses such as pneumonia to improve the detection performance on COVID-19.
\end{abstract}

\section{Introduction}

Computer-assisted diagnosis (CAD) is playing a key role in detecting COVID-19 using medical images. Deep learning models have been tailored for the detection of COVID-19. Promising exemplars include models like COVID-Net \citep{Wang2020}. One  major bottleneck of deep learning performance is the lack of data. This stops rapid model development at the onset of a novel pandemic like COVID-19, when fast and cheap screening tools are of critical importance. 

In most cases, novel illnesses have similarities to known diseases with abundant data, like COVID-19 and pneumonia. Additionally, state-of-the-art models for known diseases are being generated frequently in the medical artificial intelligence (AI) community. A low-shot approach named imprinted weights \citep{qi2018low} leverages both of these facts and can therefore be applied with little computational cost to generate high-performing models when data are scarce. These early models can jump start the novel illness classification process even with few samples, while more sophisticated models requiring more data are unavailable.

This study aims to evaluate the imprinted weights low-shot architecture, which was shown to improve the overall accuracy on all involved classes \citep{qi2018low}. Here, we adopt it for COVID-19 detection, by leveraging the abundance of pneumonia X-ray data and a pre-trained pneumonia classifier using chest radiography. Since COVID-19 is the class of interest, this study aims to evaluate the imprinted weights architecture performance on the novel minority class. Instead of overall accuracy which may be misleading for unbalanced datasets, we evaluate with per-class sensitivity and Positive predictive value (PPV) with varying n-shots. We tackle COVID-19 as a prime example of a critical novel illness classification problem with few data samples. 

\section{Literature review}
The COVID-19 outbreak attracted attention from the healthcare and medical communities. They have rapidly responded to the challenge on drug and vaccine discovery research, analyzing the impact of the disease on patients with other complications and those from high-risk groups. A temporal evolution of COVID-19 related research topics has been studied in \citet{Ebadi2021}, where the authors conducted structural modeling on COVID-19 related research, using machine learning and natural language processing techniques.

The research community in AI also played a key role in responding to the pandemic on patient screening and risk stratification. In general, the AI approaches have used handcrafted image features and those learned from deep learning. Using computed tomography (CT) scans, Shi and colleagues extracted a list of features including volume, histogram and surface features \citep{9069255}. In the ``COVID-Net" project, the authors tailored deep convolutional neural network (CNN) models for distinguishing normal, pneumonia and COVID-19 patients \citep{Wang2020}. A series of models have been trained on X-ray and CT images for screening and risk stratification \citep{Gunraj2020, Gunraj2021, wong2020covidnet}.

Due to the bottleneck of limited novel class data, low or few-shot learning techniques can greatly improve deep learning model performance \citep{10.1145/3386252}. Failing to focus on rectifying the class imbalance issue in deep learning will lead to sub-par COVID-19 screening performance. Using the low-shot imprinted weights approach \citep{qi2018low}, a preliminary study has been conducted for COVID-19 \citep{as2020covid}. In this work, we aim to  formally evaluate the imprinted weights architecture on detecting COVID-19 based on the COVIDx-CXR dataset \citep{Wang2020}.

\section{Method}
\subsection{Model Architecture}
Our model follows the architecture outlined in \citet{qi2018low} and differs in two ways from the traditional transfer learning architecture for convolutional neural networks (CNNs). After an embedding extractor, L2 normalization is applied to the embedding vector for a resulting vector of unit length. The softmax prediction layer has no bias term, and each column of its weight matrix is also normalized to unit length via L2 normalization. Our model uses ResNet-50 \citep{7780459} as the embedding extractor. Because the number of classification categories is small, we do not use a scale factor proposed in \citet{qi2018low}. The model then consists of a 256 neuron fully connected embedding layer, and a softmax classification layer. 

For comparison, a 3-class joint model is built as a baseline model. This model shares the number of neurons in the fully connected layer and the softmax classification layer, but does not have any normalization constraints.

\subsection{Implementation Details}
The embedding extractor uses ResNet-50 pre-trained on the ImageNet data set \citep{5206848}. The parameters of the fully-connected layer are initialized randomly. Input images are resized to 256x256, cropped to 224x224 and normalized. No data augmentation is used as it was described not to contribute significantly to imprinted weights \citep{qi2018low}. 

Parameter settings are as follows. Learning rate is set at $1e-3$ for pre-trained layers, and a 10x multiplier is used for the randomly initialized layers. We used exponential step decay as the learning rate reduction method during training. The step decay is applied at the same epochs for both models. The parameters of the decay are as follows: learning rate of $1e-3$, steps of 4, and decay factor of 0.94. This lr reduction method was chosen to compare the model architectures without introducing variable learning rates during training. The SGD optimizer is used with momentum of 0.9 and 1e-4 weight decay. All models are trained for 40 epochs. 

\subsection{Data Pipeline}
The COVIDx-CXR data set \citep{Wang2020} contains three image categories with train/test distributions listed in Table \ref{data-table}. We use normal and pneumonia categories to train a base classifier. We then use all COVID-19 images for inference to the base classifier to generate averaged embedding vectors used in the imprinting step. To evaluate the generalization performance of the models, 10-fold stratified cross validations are performed. All models are evaluated on the same folds. 

For low-shot analysis, the number of COVID-19 samples available for training is changed. Different number of COVID-19 samples (denoted as `n') are randomly selected from the training split after the 0.1 validation split is performed. Models with the highest overall validation accuracy are evaluated on the stratified test fold. The test fold is unseen by the model until the final evaluation. 


\begin{table}[t]
\caption{The train/test split of COVIDx-CXR data set}
\label{data-table}
\begin{center}
\begin{tabular}{llll}
\multicolumn{1}{c}{\bf COVIDx Split}  &\multicolumn{1}{c}{\bf Normal} &\multicolumn{1}{c}{\bf Pneumonia} &\multicolumn{1}{c}{\bf COVID-19}
\\ \hline \\
Train & 7966 & 5469 & 507 \\
Test & 885 & 594 & 100 \\
\end{tabular}
\end{center}
\end{table}

\subsection{Models and Configuration Variants}
\textbf{Imprinting.} To build imprinted models, a 2-class model (with the imprinted weights architecture of normalization constraints) is first trained on the normal and pneumonia data. Once the 2-class model is fine tuned, all COVID-19 images are given to the model for inference. The resulting embedded vectors are then averaged and used to set the novel class weights in a 3-class model along with the weight columns from the 2-class model. The model architectures of the 2-class and 3-class models only differ in the number of neurons in the soft-max layer.

\textbf{3-Class Joint.} The 3-class joint model uses a standard transfer learning architecture without normalization constraints. The weights of the soft-max prediction layers are randomly initialized. 

\textbf{Fine-Tuning.} In each fold, both the models are fine-tuned, and the highest validation accuracy configuration during training is selected to be evaluated on the test fold. In both cases, the models are evaluated using the same cross validation pipeline. Since the distribution of the classes is unbalanced, and only few shots of COVID-19 are used, oversampling is used such that all classes are sampled uniformly in each mini-batch.

\section{Results}

\subsection{Analysis on Number of Shots}
\begin{figure}[ht!]
\centering
\begin{subfigure}[b]{0.48\textwidth}
    \includegraphics[width=1\linewidth]{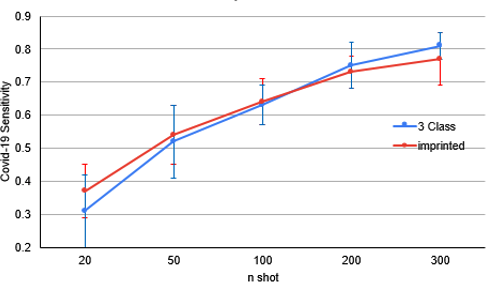}
    \caption{}
    \label{fig:n_shot_covid_sens}
\end{subfigure}
\begin{subfigure}[b]{0.48\textwidth}
    \includegraphics[width=1\linewidth]{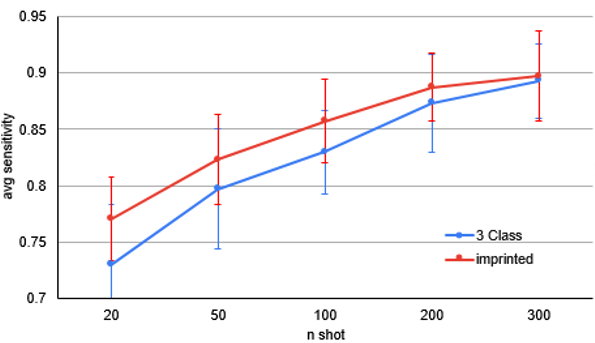}
    \caption{}
    \label{fig:n_shot_overall_sens}
\end{subfigure}
\caption[N-shot analysis of model sensitivities]{(a) COVID-19 sensitivities for Imprinted Weights and 3-class joint models as they change with increasing number of COVID-19 samples used. (b) Overall (averaged between all 3 classes) sensitivity comparison of the two models. Both (a) and (b) have their standard deviation across the folds as the error bars.}
\end{figure}

Our study focuses on the performance of the imprinted weights architecture as the number of COVID-19 samples changes. Fig. \ref{fig:n_shot_covid_sens} shows the COVID-19 sensitivity comparison between the two models. Initially at n=20, the imprinted weights model performs noticeably better. As the number of COVID-19 samples increases, this sensitivity difference decreases until at around n=100, where the average sensitivities of the two model converge. The 3-class model starts to perform better by n=300, although at this point the two models' sensitivities are very close taking the standard deviation into account. During the tests, we also noticed that imprinted model validation accuracy converged faster and with less oscillations in fewer epochs compared to the baseline model.

Fig. \ref{fig:n_shot_overall_sens} shows the averaged sensitivity from all three classes. The Imprinted Weights model outperforms the 3-class joint model more significantly, even at a high shot of n=200, before the two model sensitivities converge at n=300. The difference from Fig. \ref{fig:n_shot_covid_sens} is because the sensitivities for normal and pneumonia are consistently higher for the Imprinted Weight architecture, likely due to its weights being trained on these two classes already. 

\subsection{All COVID-19 samples}

\begin{figure}[ht!]
\centering
\begin{subfigure}[b]{0.48\textwidth}
    \includegraphics[width=0.86\linewidth]{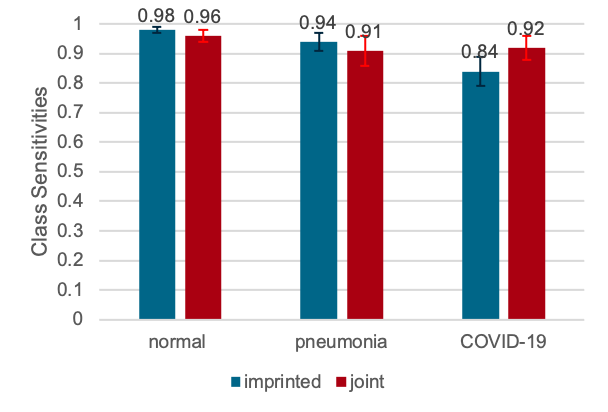}
    \caption{}
    \label{fig:all_sens}
\end{subfigure}
\begin{subfigure}[b]{0.48\textwidth}
    \includegraphics[width=1\linewidth]{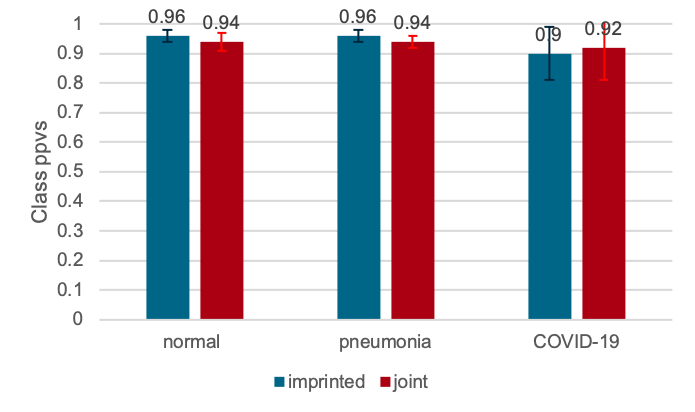}
    \caption{}
    \label{fig:all_ppv}
\end{subfigure}
\caption[Using all COVID-19 data in training]{(a) Class sensitivities for Imprinted Weights and 3-class joint models. (b) Class positive predictive value (PPV) for Imprinted Weights and 3-class joint models. Both (a) and (b) have their standard deviation across the 10 folds as the error bars.}
\label{fig:allsensppv}
\end{figure}

In addition to n-shot analysis from n=20 up to n=300, we conducted a comparison using all available COVID-19 images. Fig. \ref{fig:allsensppv} shows the per-class sensitivity and positive predictive values (PPV) for both models. The Imprinted Weights model performs better for the normal and pneumonia classes, but the 3-class joint model outperforms it slightly for the COVID-19 class. While the overall metrics of the imprinted model is higher, COVID-19 class sensitivity starts to perform worse as enough data become available. We conjecture that the Imprinted Weights model starts to under-perform past a certain threshold of data, due to its normalization constraints or the fact that COVID-19 weights are initialized based off of the base class features, which in turn caused the model to be less sensitive to COVID-19. 



\section{Discussion and Conclusion}

Through the experiments, we evaluated the effectiveness of the Imprinted Weights approach for medical data where there is one minority novel class. We also evaluated the approach more formally with 10-fold stratified cross validation, and focused on the metrics of the novel class. By applying cross validation with the same pipeline for both models, we produced more reliable results than the previous study \citep{as2020covid}. By focusing on novel class metrics instead of top-1 accuracy like \citet{qi2018low}, we found that the advantage of weight imprinting diminishes faster for COVID-19 than it does for all 3 classes. However, at low shots, weight imprinting still significantly improves COVID-19 sensitivity. Additionally, it provides smoother and faster convergence during training. At the onset of a novel pandemic with limited data, this approach can produce better models than standard transfer learning approaches with faster convergence speed. Once more data become available, more data reliant architectures can then replace this approach. 

The analysis of the imprinted weights architecture will benefit medical image classification tasks of novel illnesses with limited data. It is common for a disease to have few training samples, but have similar features to categories that the AI community has learned to classify with an abundance of images. The Imprinted Weights approach can leverage these categories to jump start the performance on classifying novel medical classes with few data. 
For future work, we plan to evaluate this approach against other few-shot architectures, and evaluate their performances as n-shots vary. 

\bibliography{iclr2021_conference}
\bibliographystyle{iclr2021_conference}


\end{document}